\documentclass[aps,prl,twocolumn,showpacs,superscriptaddress,groupedaddress]{revtex4}  % for review and submission
\usepackage{graphicx}  % needed for figures
\usepackage{dcolumn}   % needed for some tables
\usepackage{bm}        % for math
\usepackage{amssymb}   % for math
\begin{document}

% The following information is for internal review, please remove them for submission
\widetext
\leftline{Version xx as of \today}
\leftline{Primary authors: Anil Raghav}
\leftline{To be submitted to Arxiv}% (PRL, PRD-RC, PRD, PLB; choose one.)}
\leftline{Comment to {\tt raghavanil1984@gmail.com} }
%\centerline{\em D\O\ INTERNAL DOCUMENT -- NOT FOR PUBLIC DISTRIBUTION}

% the following line is for submission, including submission to the arXiv!!
%\hspace{5.2in} \mbox{Fermilab-Pub-04/xxx-E}

%\title{In-situ evidence of magnetic flux-rope, magnetic reconnection and corresponding cosmic ray (GeV) acceleration within the ICME Shock-sheath}
\title{Cosmic ray acceleration via magnetic reconnection of magnetic islands/flux-ropes }
                             
\author{Anil Raghav*}, %Ankush Bhaskar\altaffilmark{2}, Gauri Datar\altaffilmark{1} and Geeta Vichare\altaffilmark{2}}
\affiliation{University Department of Physics, University of Mumbai, Vidyanagari, Santacruz (E), Mumbai-400098, India}
\author{ Zubair Shaikh}
\affiliation{University Department of Physics, University of Mumbai, Vidyanagari, Santacruz (E), Mumbai-400098, India}
                       
\date{\today}

\begin{abstract}
The dynamic processes of magnetic reconnection and turbulence cause magnetic islands/flux-ropes generation. The in-situ observations suggest that the coalescence or/and contraction of magnetic islands are responsible to the charged  particle acceleration (keV to MeV energy range). Numerical simulations also support this acceleration mechanism.  However, the most fundamental question raise here is, does this mechanism contribute to the cosmic rays acceleration? To answer this, we report, \textit{in-situ} evidence of flux-ropes formation, their magnetic reconnection and its manifestation as cosmic ray (GeV charged particle) acceleration in interplanetary counterpart of coronal mass ejection(ICME). Further, we propose that cosmic ray (high and/or ultra-high energy) acceleration by
Fermi mechanism is valid not only through stochastic reflections of particles from the shock boundaries
but also through the boundaries of contracting magnetic islands  or/and  their merging  via magnetic re-connection.  
This has significant implications on cosmic ray origin and their acceleration process.   

%is observed via in-situ measurement at various location of interplanetary space e.g. near helio-spheric current sheets, magnetopause, Earths magnetotail etc. .  The acceleration phenomena is explained on the basis of contraction or coalescence process of magnetic islands through magnetic reconnection.
\end{abstract}

\maketitle

%\section{\label{sec:level1}First-level heading}
% sections are not used for PRL papers
\section{Introduction}

The origin of cosmic rays and its acceleration mechanism is the most fundamental problem in cosmic ray research. The relation between dimensions of particle confinement cavity and  their energy (gyro-radius) indicate cosmic ray source region in broad perspective e.g.  heliosphere (Solar (including anomalous) cosmic ray), galaxy (galactic cosmic ray), and outside galaxy (extragalactic cosmic ray) etc. 
The present understanding of cosmic ray acceleration and  the recent observations have led to a search for sources within the spacial structures e.g. the active galactic nuclei, supernovae remnants, neutron stars etc. could be the possible sources of ultra high energy galactic or extra-galactic cosmic rays.
However, how these sources contribute to cosmic ray acceleration is poorly understood. Fermi first order (diffusive shock) and second order (random collisions with interstellar clouds with characteristic velocity) acceleration process  is considered as the primal cosmic ray acceleration mechanism. Besides this, magnetic reconnection (strong localized electric field) is also thought to be contributed in cosmic ray acceleration \cite{Kulsrud 2010}.

Magnetic reconnection is a fundamental process which rearranges the magnetic field-line configuration, i.e. produces magnetic islands/structures\cite{cartwright2010heliospheric,moldwin2000small}. Further, it is responsible for energy
conversion in magnetized astrophysical and laboratory
plasmas and contributes in particle acceleration \cite{yamada2010magnetic,ren2005experimental,kronberg2004giant}.To understand the magnetic reconnection process and the corresponding particle acceleration, researchers have hunted
for magnetic structures in solar corona and space. Recent numerical simulation suggest that the magnetic islands originate from the dynamical processes of magnetic reconnection and turbulence \cite{greco2010intermittent,markidis2013kinetic,yang2015formation}. Moreover, observational studies also identified magnetic island formation regions. The literature suggest that the Sun is a natural generator of magnetic islands, e.g. CME. The regular solar wind interaction produces small scale magnetic islands as well. The occurrence of magnetic islands also observed near helio-spheric current sheets, magnetopause, Earths magnetotail etc. The magnetic reconnection process of these small-scale islands/flux-ropes give rise to an anti-reconnection electric field that can accelerate charged particles and further leads to merging or contraction of island. The particle acceleration is also possible through the contraction of islands. The trapped particles experience multiple reflections from the strongly curved field of contracting island gaining energy during each reflection via either Fermi first-order or second-order mechanism \cite{khabarova2016small,Kulsrud 2010}.% In summary, a charged particle trapped in merging or contracting magnetic island might gain several times higher energy via multiple interaction with the reconnection electric field or stochastic reflection from boundaries of closed magnetic field line structures of island.
The supporting evidences for particle acceleration (from keV to MeV energy range) are observed at various places in interplanetary space (where the magnetic islands observed) \cite{drake2006electron,chen2008observation,khabarova2015small,khabarova2016small,paschmann1979plasma, ashour2011observations}.
The recent numerical simulation based on this physical mechanism of charged particle acceleration strongly support observations \cite{drake2006electron,bian2013stochastic,zank2014particle,le2015kinetic}. However, does this physical mechanism contribute in cosmic ray acceleration processes is the fundamental question. Here we report, in-situ evidence of flux-rope formation in turbulent ICME shock-sheath, their reconnection signature and possible indication of cosmic rays (GeV energy range) acceleration. 

\section{Data}

 We have selected turbulent ICME shock-sheath \cite{venkatesan1992significance,zurbuchen2006situ,richardson2010near} which crossed the Advanced Composition Explorer (ACE)  satellite, Wind satellite and the Earth on September 24-25, 1998 for this study. To understand the spatial properties of interplanetary space  during the transit, we have used 1-minute time resolution OMNI data. The OMNI data (time corrected to the Earth bow-shock nose) includes total interplanetary magnetic field  (IMF $B_T$) and its three components ($B_x, B_y, B_z$) in GSE coordinate system, solar wind speed, plasma beta, plasma temperature and plasma density. We have also used five  minute time resolution  neutron  flux  data  taken from the NMDB database (\url{www.nmdb.eu}) to investigate cosmic rays response to this ICME shock-sheath transit. The cosmic ray data processing method used here are briefly discussed in Raghav et al (2016) \cite{raghav2016does}. Generally,  A 2D-hodogram analysis is widely used in magnetospheric physics and express impressive visualization of rotating IMF within the magnetic island events \cite{khabarova2015small,khabarova2016small}. The observation of semicircle/circle arc in one of the planes $B_x-B_y$ or $B_z-B_y$ or $B_z-B_x$ during island crossing manifest as rotation of IMF. However, we will not have information of time evolution of rotating plane in 2D-hodogram method. Therefore, 4D-hodogram method is performed, in which 1 second time resolution ACE satellite data ($B_x$, $B_y$ \& $B_z$ in GSE coordinate system) is used. Beside this, to cross verify the signature of magnetic reconnection of flux-ropes electron density, velocity x-component (in GSE-coordinate system), electron flux (5 kev and 20 keV) and ion flux (0.14 keV, 4 keV and 19 keV) data  from Wind satellite are used. 
 \section{Observations and discussion} 
 
\begin{figure}
 	%  \centering
 	%\includegraphics[width=18 cm height = 26cm ]{longi_stack_nov04_cr.eps}
 	\includegraphics[width=0.5\textwidth]{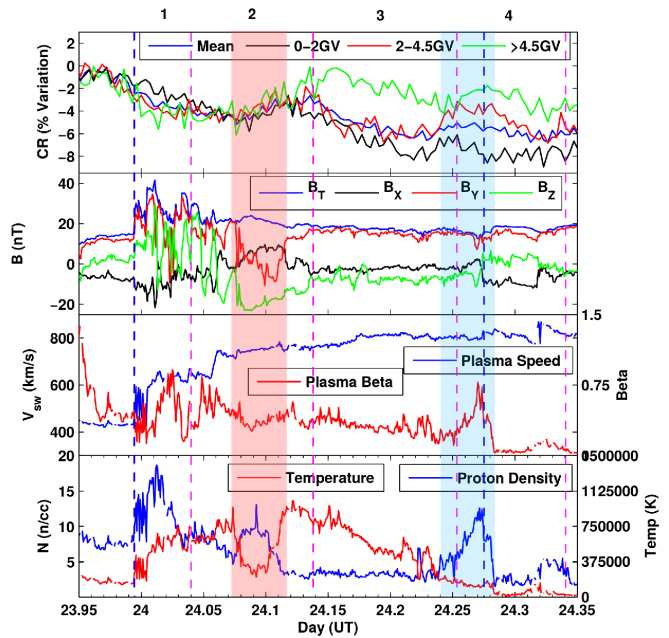}
 	\caption{\textit{ICME crossing event occurred on September 24-25, 1998. The figure has four panels, top most panel shows temporal variation of normalized neutron flux with their respective band of rigidities. The $2^{nd}$ panel show interplanetary magnetic field ($B_{Total}$ and $B_X$,$B_Y$,$B_Z$-component). The $3^{rd}$ panel from top illustrates solar wind speed and plasma beta variations. The bottom panel depicts proton density and plasma temperature variation.}} % The boundaries of shock-sheath region are defined using dash blue vertical lines. The dash magenta vertical lines divide shock sheath in four different regions. The red-brown semi-transparent rectangle shows signature of rotational magnetic island, however the blue semi-transparent rectangle shows signature of possible reconnection and cosmic ray acceleration.}}
 	\label{fig:1}
 \end{figure}

\begin{figure}
 	%  \centering
 	%\includegraphics[width=18 cm height = 26cm ]{longi_stack_nov04_cr.eps}
 	\includegraphics[width=0.5\textwidth]{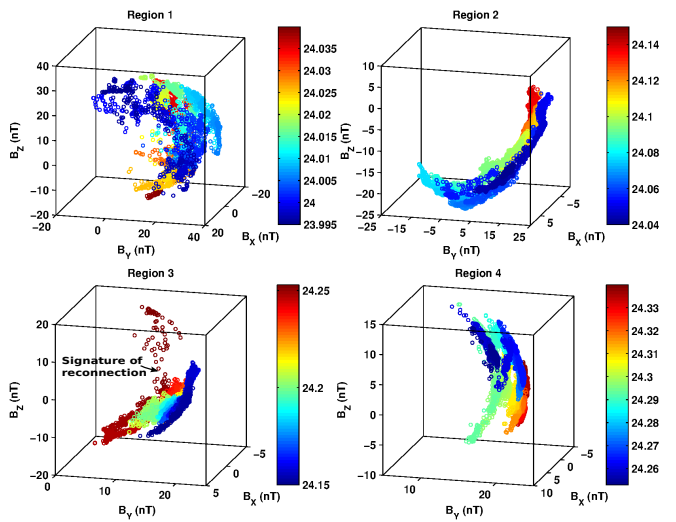}
 	\caption{\textit{The four 4D hodograms for selected ICME shock sheath region. Each hodogram shows $B_z - B_y$ as front plane of 3D projection and $B_x$ variations are shown as in-out directions. The  temporal variations of event crossing is shown as color-bar. The top left and right panel hodogram illustrate region 1 and region 2 and the bottom left and right panel hodograms depict region 3 and region 4 of Figure~\ref{fig:1} respectively. }}
 	\label{fig:2}
 \end{figure}

 The variations of comic ray flux and interplanetary parameter  during the selected event is shown in Figure (1). The ICME boundaries are defined using Richardson \& Cane (2010) \cite{richardson2010near} and presented as dashed blue vertical lines. The ICME shock-sheath region is divided in 4 parts for better investigation using dashed magenta vertical lines. Figure \ref{fig:2} shows 4 hodograms for all selected regions of the Figure \ref{fig:1}.
 
 The sharp enhancement in total IMF $B_T$ and solar wind suggest the onset of interplanetary shock-front at the Earth's Bow-shock nose. In region 1, random fluctuations are seen in total IMF including all its components. We have also observed  sudden increase in plasma temperature and density.
 Moreover, the top left panel hodogram (for region 1) of Figure~\ref{fig:2}, shows fuzzy signature of semicircle but highly dominated by random fluctuations. These observations could be ascribed as the residence of turbulence in shock-front.  Compression and plasma heating in shock-front could be the cause of this observed turbulence. 

In region 2, especially the faint red-brown shaded part demonstrates clear rotation in $B_z$ and $B_y$ components of IMF, however total IMF is gradually increasing and then decreasing. The plasma temperature/density shows decrease/increase, and solar wind shows steady variations during the transit of red-brown shaded region. The top right panel hodogram (for region 2) in Figure~\ref{fig:2}, shows explicit visualization of semicircle with varying circle arc. This structure further continued in remaining two hodograms (bottom left and right). These are clear evidences of magnetic island/flux-rope formation in shock sheath. It is important to note that the red-brown shaded region is the only possible signature of flux-rope is seen from Figure \ref{fig:1}. However, 4D hodogram depicts much better visualization of the flux-rope forming regions and extend its boundaries to region 3.
 \begin{figure}
 	%  \centering
 	%\includegraphics[width=18 cm height = 26cm ]{longi_stack_nov04_cr.eps}
 	\includegraphics[width=0.5\textwidth]{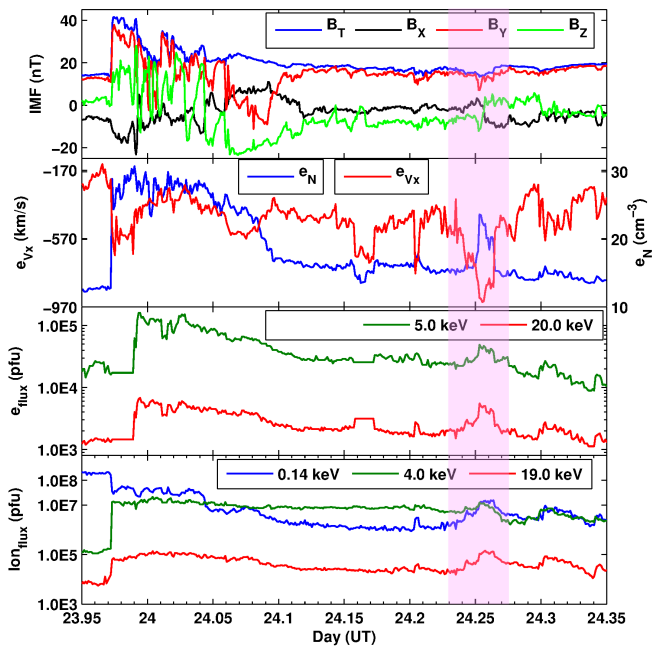}
 	\caption{\textit{IMF and Particle flux variation during ICME transit occurred on September 25, 1998. The figure has four panels, top most panel shows temporal variation of interplanetary magnetic field ($B_{Total}$ and $B_X$,$B_Y$,$B_Z$-component). The $2^{nd}$ panel show electron density and its velocity component in x direction. The $3^{rd}$ panel from top illustrates 5 keV and 20 keV electron flux variation. The bottom panel depicts 0.14 keV, 4 keV and 19 keV ion flux variations. }}
 	\label{fig:5}
 \end{figure}
During region 3 transit, IMF B and its components remain steady and IMF was mainly directed in Y-direction i.e dawn to
dusk. The solar wind speed, plasma beta and plasma density show steady variations, only plasma temperature gradually decreases to ambient value. In region 4 specially blue shaded region shows sharp transition  in all IMF components. The plasma density, plasma beta and cosmic ray flux depict corresponding enhancement. Basically, flux-rope of the shock-sheath region and ICME flux-rope boundary lie in this region. The left bottom hodogram (for region 3), shows sharp switch in the oscillating orientation (check brown circle data) which further continued in region 4.  We have also noted similar oscillation tripping at the onset of ICME flux-rope (in bottom right hodogram for region 4). This could be the indication of combination of two different flux-ropes  (shock-sheath and ICME) via magnetic reconnection. To support the observed signature of magnetic reconnection electron and ion flux data are investigated. Figure \ref{fig:5} shows temporal variation of electron density, its velocity in x direction, electron and ion flux from wind satellite. Moreover, all studied parameters (except IMF which used for reference between Figure \ref{fig:1} and \ref{fig:5}) explicitly depicts enhancement in pink shaded region of Figure \ref{fig:5}. These observations are clear indication of magnetic reconnection and corresponding charged particle acceleration.

\begin{figure}
	%  \centering
	%\includegraphics[width=18 cm height = 26cm ]{longi_stack_nov04_cr.eps}
	\includegraphics[width=0.5\textwidth]{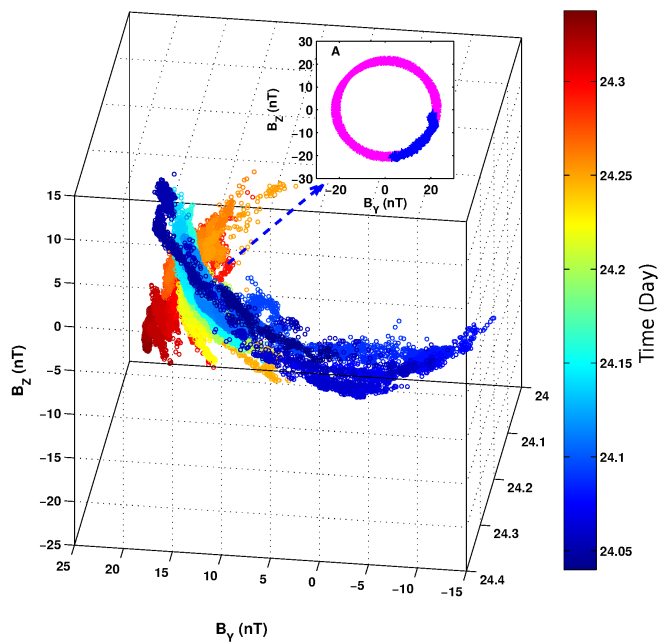}
	\caption{\textit{The combined 4D hodograms for region 2 to 4 as shown in Figure ~\ref{fig:1}. The hodogram shows $B_z - B_y$ as front plane of 3D projection and temporal variations are shown as in-out directions as well as color-bar. The sub-figure (A) shows the blue top layer of semicircle of 4D hodogarm in  $B_z - B_y$ plane. The magenta circle is simulated by assuming [$B_Z = 23*sin(2\pi t/T)* rand() $] and [$B_Y = 23*cos(2\pi t/T)* rand() $].}}
	\label{fig:3}
\end{figure}

\begin{figure*}
	%  \centering
	%\includegraphics[width=18 cm height = 26cm ]{longi_stack_nov04_cr.eps}
	\includegraphics[width=1\textwidth]{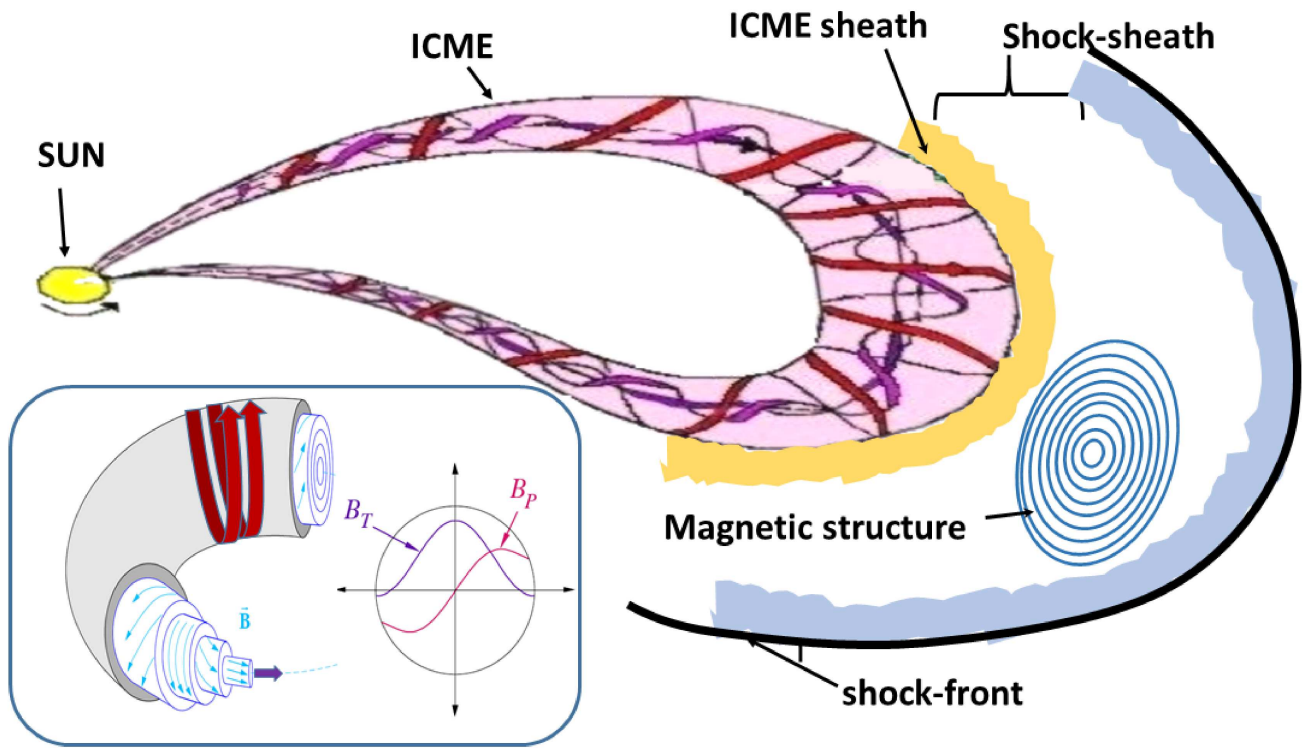}
	\caption {\textit{The updated schematic diagram (artistic illustration) of ICME in interplanetary space.  }}
	\label{fig:4}
\end{figure*}

To understand origin and evolution of observed flux-rope within shock-sheath, the complete overview of magnetic flux-ropes formed in shock-sheath region (from region 2 to 4 in Figure~\ref{fig:1})) is presented as 4D hodogram in Figure~\ref{fig:3}. Initially, the arc length of circle is decreasing to minimum and then again increasing with slowly continuous oscillation of the arc-circle plane (see the circle arc with different color, on-line only). %Couple of times,we have also observed sharp orientation flipping in oscillating plane (see blue, red-yellow and red circle arc). This could be interpreted as the signature of magnetic reconnection.
Moreover, to estimate the phase difference between $B_Z$ and $B_Y$, the top circle arc layer is selected and shown as the sub-figure (A) in Figure~\ref{fig:3}. The IMF components ($B_Y ~\&~ B_Z$) are simulated using equations as [$B_Z = 23*sin(2\pi t/1600)* rand() $] and [$B_Y = 22*cos(2\pi t/1600)* rand()$]. Here, $t$ varies from 1 to 1600 and $rand()$ is computer generated random number which varies from 0 to 1. The simulated data is presented as magenta circle in sub-figure (A).  The simulated data is clearly fitted with the observed top circle arc layer data. This proves that the $B_Z$ and $B_Y$ are having phase angle $\pi /2$. Further, in region 2 of figure~\ref{fig:1}, we have observed $B_Z$ and $B_Y$ are rotating and finally total IMF is oriented along the y-direction. Now, if we assume toroid shape of flux-rope, in which $B_Z$ and  $B_Y$ presents poloidal and toroidal field direction respectively as shown in Figure~\ref{fig:4} and the ACE spacecraft crossing direction is not parallel to the axis of the flux-rope. Then observations suggest that the different circle arc observed in hodogram are the different layers of flux-rope. From top to inside, at every layer, the poloidal component of magnetic field slowly aligning with the toroidal component. This can be seen as the oscillation of arc plane in Figure~\ref{fig:3}. Finally $B_Z$ poloidal component reach to its minimum whereas the total IMF of flux-rope is represented by toroidal component i.e. $B_Y$. This ascribed that the plasma in shock-sheath region minimize their potential energy. This physical process is known as plasma relaxation (self-organization of a plasma) by magnetic reconnection \cite{taylor1986relaxation}. Figure~\ref{fig:4} (artistic illustration) exhibits the updated visualization of ICME propagation in interplanetary space. The ICME flux-rope with turbulent shock-sheath is the earlier hypothesis. In this work, we advances this traditional hypothesis and demonstrate the first in-situ evidence of flux-rope formation other than ICME flux-rope in shock-sheath region.  The flux-rope could be originated from the interaction of ICME with ambient solar wind or fragmentation of ICME flux-rope due to solar wind interaction \cite{khabarova2016small}.

%\twocolumn

Beside this, the enhancement of cosmic rays flux (in all observed energy band) is unambiguously observed in red and blue shaded region of Figure~\ref{fig:1}. In addition, in Figure \ref{fig:2}, the top right hodogram shows the indication of rotating magnetic island formation and bottom left hodogram demonstrates the signature of magnetic reconnection during respective time. The enhancement in red shaded region could be understood on the basis of gradual decrease in toroidal and poloidal field strength of flux-rope during corresponding period. Moreover, the enhancement in charged particle flux is generally interpreted as the particle energization i.e. acceleration \cite{khabarova2015small}. The simulation studies suggest that the particle acceleration is possible when particle gyro-radius is smaller than the magnetic islands scale length \cite{le2015energetic}. That means the accelerated particle energy range is depends on the dimensions of magnetic island. In present case, the dimension of observed magnetic island is estimated approximately $3 * 10^7 ~km$ using the time duration of magnetic island crossing and average solar wind speed. The estimated dimension is much higher than the order of cosmic ray (1-10 GeV energy  range observed by neutron monitors) gyro-radius. %In given physical scenario, the entered cosmic ray should trap inside the observed magnetic flux-rope and gyrate.
Therefore, the enhancement in cosmic ray flux in red shaded region also could be the outcome of acceleration through a stochastic repeated reflections (first-order Fermi mechanism (in the case of compressible contraction) or a second-order Fermi mechanism (if the contraction is incompressible)) from contracting or merging flux-rope \cite{khabarova2015small,khabarova2016small}. 
However, enhancement in cosmic ray flux in blue shaded region could be ascribed as explicit evidence of cosmic ray acceleration via magnetic reconnection of shock-sheath flux-rope and ICME flux-rope.

\section{Conclusions}

A charged particle acceleration via contraction and/or merging of magnetic islands through magnetic reconnection is observationally evident in keV to MeV energy range and strongly supported by numerical simulations. The present work extend this accelerating process to GeV energy range charged particles i.e. cosmic rays. On the basis of dimensional argument including field strength inside the magnetic island and the particle energy, we propose that the same processes should contribute significantly in high energy and/or ultra-high energy cosmic ray acceleration. This further imply that not only the active core region but  also transient disturbances in the spatially extended region of active core-region are responsible for cosmic ray acceleration. For example, %in supernovae explosion, cosmic rays are trapped in expanding supernova bubble and lost their energy adiabatically by decompression. Therefore, it is impossible for the cosmic rays to come directly from the remnant or from a pulsar arising at the center of explosion. However, 
after supernova expansion, the hot thermal material generates shocks into the undisturbed interstellar medium \cite{Kulsrud 2010}. This physical scenario is similar to the ICME shock in heliosphere. The transient disturbances could give rise to turbulent and dynamic conditions leads to magnetic islands formation. Moreover, the interactions of these magnetic structures through magnetic reconnection induce merging and contraction of islands, ultimately contributes to the charge particle (cosmic ray) acceleration. % This further supported by fact that the young supernovae (small bubble) or older supernovae (with large bubbles with weaker shock i.e. transients) should not contribute in cosmic rays acceleration. Moreover, Intermediate age supernovae (moderate size bubble with active strong transients) significantly contributes in cosmic ray acceleration \cite{Kulsrud 2010}.

In summary, we conclude that the cosmic rays acceleration by Fermi mechanism is valid not only through stochastic reflections of particles from the shock boundaries but also through merging and/or contraction of magnetic islands via magnetic reconnection. This may provide some  insight into the origin of ultra-high energy cosmic rays. The ultra-high energy cosmic rays may not be one-step process but a multi-step one. The region responsible for final acceleration may be thought of as the origin of these ultra-high energy cosmic rays. .

\subsection{Acknowledgment}
We acknowledge the NMDB database (www.nmdb.eu) founded under the European Union's FP7 programme (contract no. 213007). We are also thankful to all neutron monitor observatories listed on website. We are thankful to CDAWeb and ACE science center for making interplanetary data available. We are thankful to Department of Physics (Autonomous), University of Mumbai, for providing us facilities for fulfillment of this work. Authors would also like to thank A. Bhaskar, S. Kasthurirangan, N. Bijewar and M. Gadyali for valuable discussion.

%% Here is the endmatter stuff: Supplementary Info, etc.
%% Use \item's to separate, default label is "Acknowledgements"

\end{document}